%% file: main.tex
\documentclass{jfm}

\graphicspath{{figures/}} 
\usepackage{graphicx} 
\usepackage{natbib} 
\usepackage{hyperref} 
\usepackage{newtxtext,newtxmath}
\usepackage{bm} 
\usepackage{amssymb}
\usepackage{ar}
\usepackage[]{lineno}

\newcommand{\aver}[1]{ \! \left\langle {#1} \right \rangle \!}

\usepackage[dvipsnames]{xcolor}

\title[]{Passive scalar cascade in the intermediate layer of 
  turbulent channel flow for $Pr\leq 1$}

\author[]{Emanuele Gallorini\aff{1}\corresp{\email{emanuele.gallorini@polimi.it}}, Shingo Motoki\aff{2}, Genta
  Kawahara\aff{2} and Christos Vassilicos \aff{1}} \affiliation{

\aff{1} Univ. Lille, CNRS, ONERA, Arts et Metiers Institute of
Technology, Centrale Lille, UMR 9014 - LMFL - Laboratoire de Mécanique
des Fluides de Lille - Kampé de Fériet, F-59000 Lille, France
\aff{2} Graduate School of Engineering Science, University of Osaka, 1-3
Machikaneyama, Toyonaka, Osaka 560-8531, Japan } 
\begin{document}

\maketitle

\begin{abstract} 

Similarities and differences between Kolmogorov scale-by-scale
equilibria/non-equilibria for velocity and scalar fields are
investigated in the intermediate layer of a fully developed turbulent
channel flow with a passive scalar/temperature field driven by a
uniform heat source. The analysis is based on intermediate asymptotics
and direct numerical simulations at different Prandtl numbers lower
than unity. Similarly to what happens to the velocity fluctuations,
for the fluctuating scalar field Kolmogorov scale-by-scale equilibrium
is achieved asymptotically around a length scale $r_{min}$, which is
located below the inertial range. The lengthscale $r_{min}$ and the
ratio between the inter-scale transfer and dissipation rates at
$r_{min}$ vary following power laws of the Prandtl number, with
exponents determined by matched asymptotics based on the hypothesis of
homogeneous two-point physics in non-homogeneous turbulence. The
interscale transfer rates of turbulent kinetic energy and passive
scalar variance are globally similar but show evident differences when
their aligned/anti-aligned contributions are considered.

\end{abstract} 


\input{./tex/Intro.tex}

\input{./tex/Methods.tex}

\input{./tex/Results.tex}
\input{./tex/Conclusions.tex}


\section*{Declaration of Interests} The authors report no conflict of interest.

\section*{Acknowledgments} 
J.C.V. acknowledges funding by the European Union (ERC Advanced Grant
NoStaHo, project number 101054117). Views and opinions expressed are,
however, those of the author(s) only and do not necessarily reflect
those of the European Union or the European Research Council. Neither
the European Union nor the granting authority can be held responsible
for them. 
This project was provided with computing HPC and storage resources by GENCI at CINES thanks to the grant 2025-A0182A01741 on the supercomputer Adastra's GENOA partition .

\bibliographystyle{jfm} 
\bibliography{./Wallturb}

\end{document}

%% file: tex/Intro.tex
\section{Introduction}

The study of passive scalars representing a dilute diffusive
contaminant in a fluid flow or a temperature field with small
variations at low Mach numbers, has applications in mixing,
combustion, and pollution. It presents complexities that elevate it
above being a simple corollary of the fluid flow
\citep{warhaft-2000}. Pressure-driven turbulent channel flow (TCF)
between two parallel plane walls is a standard test case for
turbulence, and this remains true when a passive scalar is present.
Several Direct Numerical Simulation (DNS) studies of TCF
\citep{abe-kawamura-matsuo-2004,antonia-abe-kawamura-2009,
  pirozzoli-bernardini-orlandi-2016} have highlighted similarities
between velocity and passive scalar fields. Despite minor differences
relating to forcing and computational setup, these works have shown
that both the mean flow and mean scalar profiles have a region of
logarithmic dependence on wall normal direction. From the energetic
point of view, this region (intermediate layer) is characterized by an
approximate balance between turbulence production and dissipation
rates for both the turbulent kinetic energy and the passive scalar
variance \citep{pope-2000}. However, the picture appears more
complicated when the way turbulent kinetic energy is distributed and
transported among different scales and locations is investigated
\citep{cimarelli-deangelis-casciola-2013,
  gatti-etal-2020}. Nevertheless, the matched asymptotic analysis of
the scale-by-scale turbulent kinetic energy budget in the intermediate
layer of a TCF by \cite{apostolidis-laval-vassilicos-2023} has shown
that a Kolmogorov-type equilibrium between turbulence dissipation and
interscale transfer rate is achieved asymptotically (with increasing
Reynolds number) only around the Taylor length, and therefore not in
the inertial range. These conclusions for a case of a stationary
non-homogeneous turbulence agree with the results obtained for
non-stationary (freely decaying) homogeneous isotropic turbulence
(HIT) far from initial conditions by \cite{lundgren-2002}, whose
matched asymptotic analysis of the K\'arm\'an-Howarth equation leads
to the conclusion that the interscale turbulence transfer rate has an
extremum at a length scale $r_{min}$ proportional to the Taylor length
$\lambda$. Wind tunnel data of freely decaying HIT
\citep{obligado-vassilicos-2019} yield $r_{min} \simeq 1.5\lambda$,
and \cite{meldi-vassilicos-2021} found $r_{min} \simeq 1.12\lambda$
for Taylor length-based Reynolds numbers $Re_\lambda = 10^2$ to $10^6$
using an eddy-damped quasinormal Markovian (EDQNM) model of decaying
HIT. These works support the idea that a class of turbulent flows
exist (including both non-stationary homogeneous and stationary
non-homogeneous turbulence) where Kolmogorov-like equilibrium is
asymptotically achieved solely in the vicinity of the Taylor length
rather than within an inertial range.

But does such a type of localised scale-by-scale budget also hold for
a passive scalar field, and if so, around what length-scale? And with
what similarities to and differences from the interscale transfer
properties of the supporting velocity field? In this paper, we address
these questions by studying the interscale transfer rate of passive
scalar variance in a TCF, limiting the analysis to the case when the
diffusivity of the passive scalar is higher than that of the velocity
(i.e., Prandtl number $Pr \le 1$). Our methods are described in section
\ref{sec:methods}, our results in section \ref{sec:results}, and we
conclude in section \ref{sec:conclusions}.

%% file: tex/Methods.tex
\section{Methods}
\label{sec:methods}
The present work focuses on turbulent heat and momentum transfer in
internally heated channel flow between isothermal impermeable
walls. The coordinates $x$, $y$, and $z$ represent streamwise,
wall-normal and spanwise directions, $\bm{u}(\bm{x},t)=(u,v,w)$ is the
velocity vector with components in these respective directions, and
$\theta(\bm{x},t)$ is the passive temperature field. We use the method
of matched asymptotic expansions \citep{vanDyke-1964}
and data from DNS of TCF. The channel has dimensions $L_x\times L_y
\times L_z = 2\pi h\times 2h \times \pi h$ . The governing equations are the incompressible
Navier-Stokes equations for the velocity field and the energy equation, i.e.
advection-diffusion equation for the temperature with a uniform
internal heat source to maintain constant bulk mean temperature $T_b$.
At the walls ($y=0$ and $2h$), no-slip boundary conditions apply for
the velocity, and a similar Dirichlet condition applies for the
temperature. The streamwise and spanwise directions are periodic. The
flow is driven by a variable pressure gradient which is adjusted at each 
timestep to maintain a constant  flow rate. In the following,
$\nu$ and $\kappa$ are the kinematic viscosity and thermal diffusivity
respectively. The flow is characterized by two dimensionless
parameters: the bulk Reynolds number $Re_b=2hU_b/\nu$ and the Prandtl
number $Pr=\nu/\kappa$. The present DNS data analysis investigates
several cases at $Re_b=40000$ with different Prandtl numbers $Pr=1$,
$Pr=0.75$, $Pr=0.5$, and $Pr=0.25$.  The friction Reynolds number,
based on the friction velocity $u_\tau=\sqrt{\overline{\tau}_w/\rho}$
($\overline{\tau}_w$ is the mean wall shear stress) is
$Re_\tau=1000$. The friction temperature is defined as
$\theta_\tau=\kappa\text{d}T/\text{d}y|_{w}/u_\tau$, where
$\text{d}T/\text{d}y|_{w}$ is the wall normal derivative of the mean
temperature $T$ at the $y=0$ wall. The details of the computational
procedures and the generation of the database can be found in
\cite{motoki-etal-2022}. The cases with $Pr\neq 1$ are absent in the
original database; therefore, additional simulations have been
performed for the same configuration using the pseudo-spectral code developed by
\cite{luchini-quadrio-2006}.  The domain is discretised with
$n_x\times n_z = 512 \times 512$ Fourier modes and $n_y=515$ points in
the wall-normal direction, spaced according to a hyperbolic-tangent
distribution. The resulting resolution is $\Delta x^+=8.2$ and $\Delta
z^+=4.1$ after dealiasing, while $\Delta y^+\approx 1$ at the wall and
$\Delta y^+\approx 6.7$ at the centerline, equivalent to similar works
on turbulent channel flow at this Reynolds number
\citep{gatti-etal-2020}, where the superscripts $+$ stand for the normalised length with $\delta_\nu\equiv \nu/u_\tau$. The limited extension of the domain could
affect the dynamics of very large-scale structures. However, we expect
a limited impact at the scales that we focus on in this study, as
confirmed by the agreement between the theoretical predictions and the
numerical results in the following section.

We extend the matched asymptotic expansions approach of
\cite{apostolidis-laval-vassilicos-2023} to passive scalar interscale
transfers. They applied this approach to the
K\'{a}rm\'{a}n-Howarth-Monin-Hill (KHMH) equation \citep{hill-2001}
for the turbulent velocity field and we apply it here to the fully
generalised Yaglom equation for the passive scalar \citep{danaila-etal-1999} 
obtained by \cite{hillB-2002}.
These are the budget equations for the second-order structure
functions of the velocity and passive scalar respectively. Decomposing
the velocity and temperature fields into mean ($\bm{U},T$) and
fluctuating ($\bm{u}',\theta'$) fields, these structure functions are
$\aver{\delta \bm{u}'^2}$ and $\aver{\delta \theta'^2}$, where the
brackets signify an average; $\delta \bm{u}' \equiv
\bm{u}'(\bm{x}+\bm{r}/2,t)-\bm{u}'(\bm{x}-\bm{r}/2,t)$ and $\delta
\theta' \equiv
\theta'(\bm{x}+\bm{r}/2,t)-\theta'(\bm{x}-\bm{r}/2,t)$. The centroid
$\bm{x}$ (components $x_1 \equiv x$, $x_2 \equiv y$, $x_3 \equiv z$)
determines spatial location and the separation vector $\bm{r}$
(components $r_i$, $i=1,2,3$) determines length scales. We introduce
the additional notation
$\bm{u}'^*\equiv\frac{\bm{u'}(\bm{x}+\bm{r}/2,t) +
  \bm{u'}(\bm{x}-\bm{r}/2,t) }{2}$ with obvious extension to all other
fields, in particular the mean velocity field.

The KHMH and fully generalised Yaglom equations are derived directly
from the Navier-Stokes and the advection-diffusion equations
\citep{hill-2001, hillB-2002}. For the first, we refer the interested
reader to \cite{apostolidis-laval-vassilicos-2023}. The latter reads:

\begin{gather}
\underbrace{\frac{\partial \aver{\delta \theta'^2}}{\partial
    t}}_{A_{Tt}} + \underbrace{\frac{\partial U_i^*\aver{\delta \theta'^2}}{\partial x_i}}_{\mathcal{T}_{Tm}}+
\underbrace{\frac{\partial\aver{ u_i'^*\delta
      \theta'^2}}{\partial x_i}}_{\mathcal{T}_{T}}+ \underbrace{\frac{\partial
    \delta U_i\aver{\delta \theta'^2}}{\partial
    r_i}}_{\varPi_{Tm}}+ \underbrace{\frac{\partial\aver{ \delta
      u_i'\delta \theta'^2}}{\partial r_i}}_{\varPi_{T}} =
\nonumber \\ \underbrace{-2\aver{ u_i'^* 
    \delta \theta'}\frac{\partial\delta T}{\partial x_i}
  -2\aver{\delta u_i' \delta \theta'}\frac{\partial\delta
    T}{\partial r_i} }_{\mathcal{P}_T} +
\underbrace{\frac{\kappa}{2}\frac{\partial^2\aver{
      \delta \theta'^2}}{\partial x_i^2}}_{D_{Tx}}+
\underbrace{2\kappa\frac{\partial^2\aver{\delta
      \theta'^2}}{\partial
    r_i^2}}_{D_{Tr}}-\underbrace{2(\varepsilon_T^p+\varepsilon_T^m)}_{\varepsilon_T}.
\label{eq:KHMHSTOT}
\end{gather}
In these equations,
$\varepsilon_T^p\equiv\kappa\aver{\partial \theta'^p/\partial
  x_j^p\partial \theta'^p/\partial x_j^p}$,
$\varepsilon_T^m\equiv\kappa\aver{\partial \theta'^m/\partial
  x_j^m\partial \theta'^m/\partial x_j^m}$ where the $p$ and $m$
superscripts indicate quantities evaluated at $\bm{x}+\bm{r}/2$ and
$\bm{x}-\bm{r}/2$ respectively.  The different terms of equation
\ref{eq:KHMHSTOT} represent different mechanisms: $A_{Tt}$ represent
redistribution due to unsteadiness, $\mathcal{T}_{Tm}$ and
$\mathcal{T}_{T}$ are transports in physical space due to mean and
fluctuating velocity, $\varPi_{Tm}$ and $\varPi_{T}$ are interscale
transfer rates due to mean and fluctuating flow, $\mathcal{P}_T$ is
the production rate, $D_{Tx}$ and $D_{Tr}$ are diffusive terms in
physical and scale spaces, and $\varepsilon_T$ is the turbulent scalar
dissipation rate.  This work focuses on $Pr \le 1$.

%% file: tex/Results.tex
\section{Results}
\label{sec:results}

The KHMH equation and the generalized Yaglom equation
(\ref{eq:KHMHSTOT}) can be simplified by tailoring them to the channel
flow for which the mean flow is non-zero only in the streamwise
direction. Averages are performed in time and along the
homogeneous/periodic streamwise and spanwise directions. Hence,
$A_{Tt}$ and $\mathcal{T}_{Tm}$ vanish and the remaining terms of the budget
depend on $(y,\bm{r})$. The presence of an approximately logarithmic
mean flow in the intermediate layer for $Re_{\tau}\gg 1$
\citep{motoki-etal-2022} implies that
$\varPi_{Tm}$ is close to zero in the intermediate layer $\delta_{\nu}
\ll y \ll h$ if $r_2 \ll 2y$. This follows in the same way that
\cite{apostolidis-laval-vassilicos-2023} showed the equivalent
($\varPi_m$) term in the KHMH equation to be about zero in this layer,
as $\delta U_{2} = \delta U_{3} = 0$ and $\delta U_1 = (u_\tau/k)\ln
[(1+r_2/y)(1-r_2/y)] \approx 0$ for $r_{2}/y \ll 1$ ($k$ is the von
K\'arm\'an coefficient). These authors also showed that the
inter-space transfer rate $\mathcal{T}$ and the pressure-velocity term
$\mathcal{T}_p$ are also negligible in the intermediate layer. Under
the same hypothesis of well-mixed turbulence in this layer when
$Re_{\tau}\gg 1$ and $u_{\tau}h/\kappa =Re_{\tau}Pr\gg 1$, we assume
that the inter-space transfer rate $\mathcal{T}_T$ is also negligible
in this intermediate layer. The low magnitude of $\varXi_t \equiv
\varPi_{Tm}+\mathcal{T}_T$ compared to the remaining terms of equation
\ref{eq:KHMHSTOT} is confirmed with DNS data in sub-section
\ref{sec:verification}. We therefore have the following simplified
balance where every term has been averaged over a sphere in ${\bm r}$
space, e.g.  $\varPi^{v}\equiv (6/\pi r^{3})\int_{S(r)} \varPi
\,\text{d}^{3}{\bm r}$ where $S(r)$ is the sphere of diameter $r\equiv
\vert{\bm r}\vert$:
\begin{equation}
\varPi_{T}^v \approx  \mathcal{P}_T^v+D_{Tx}^v+D_{Tr}^v-\varepsilon_T^v
\label{eq:budgt}
\end{equation}
valid for $r \ll 2y$ in the intermediate layer
$\delta_\nu \ll y \ll h$.
Note that the pressure enters the KHMH equation for the two-point
turbulent kinetic energy through a transport term which is negligible
in the intermediate layer \citep{apostolidis-laval-vassilicos-2023} so
that this KHMH equation features the same qualitative terms as
(\ref{eq:budgt}) in that layer (inter-scale transfer, production,
viscous diffusion and dissipation).

\subsection{Outer and inner asymptotics}
\label{sec:asym}

The spherically averaged scalar inter-scale transfer rate is:
\begin{equation}
\varPi^v_{T} = \frac{3}{2\pi}\int \aver{ \frac{\delta \bm{u}'\cdot
    \hat{\bm{r}}}{r} \delta \theta'\delta \theta'} \, \text{d}\Omega_r
\equiv \frac{S_{T3}(r,y)}{r}
\end{equation}
after use of the Gauss divergence theorem ($\Omega_r$ is the solid
angle in $\bm{r}$ space and $\hat{\bm{r}} \equiv \bm{r}/r$). The
logarithmic law $\text{d}T/\text{d}y=\theta_\tau /(k_{\theta}y) $ (where $k_\theta$ is
a dimensionless coefficient) in the intermediate layer for $Re_{\tau}$
and $Re_{\tau} Pr$ much larger than 1 \citep{kader-yaglom-1972} has
been confirmed by \citep{motoki-etal-2022} in their DNS of TCF and
implies that the production term takes the form:
\begin{equation}
\mathcal{P}^v_{T} \approx - \frac{6}{\pi
  r^3}\frac{\theta_\tau^2u_\tau}{k_\theta y}\int_0^{r/2}\rho^2
\Bigl[\frac{S_{T12}(\rho,y)}{u_\tau\theta_\tau} -
  \frac{S_{T1\times2}(\rho,y)}{u_\tau\theta_\tau}\Bigr] \,
\text{d}\rho
\end{equation}
where:
\begin{gather}
S_{T12}(\rho,y) = 2\int \aver{\delta v' \delta \theta'} \Bigl[
  1-\Bigl(\frac{r_y}{2y} \Bigr)^2 \Bigr]^{-1}\, \text{d}\Omega_r
\\ S_{T1\times2}(\rho,y) = 2 \int \aver{v'^* \delta \theta'}
\frac{r_y}{y}\Bigl[ 1-\Bigl(\frac{r_y}{2y} \Bigr)^2 \Bigr]^{-1}\,
\text{d}\Omega_r .
\end{gather}
Note, however, that our hypothesis that a well-developed logarithmic
region exists for $T$ is not supported by DNS data when $Pr$ is low
enough \citep{alcantara-avila-etal-2018}.

Finally, the molecular diffusion rate can be expressed in terms of
$S_{T2} = \int \aver{\delta \theta' \delta \theta'} \,
\text{d}\Omega_r$ as follows:
\begin{equation}
D^v_{Tx}+D^v_{Tr} = \frac{3\kappa}{\pi r^3} \int_0^{r/2} \rho^2
\frac{\text{d}^2S_{T2}}{\text{d}y^2}(\rho,y) \, \text{d} \rho +
\frac{3\kappa}{\pi r}\frac{\text{d}S_{T2}}{\text{d}r}(r,y).
\end{equation} 
According to our DNS data, $\vert S_{T1\times2}\vert \ll \vert
S_{T12}\vert$, as might be expected. We therefore neglect
$S_{T1\times2}$, and the scale-by-scale budget (\ref{eq:budgt}) can be
fully described in terms of the integrals $S_{T2}$, $S_{T3}$,
$S_{T12}$ and an expression for the scalar dissipation rate
$\varepsilon^v_T$.

At this point we invoke the hypothesis of homogeneous two-point
physics in non-homogeneous turbulence already introduced by
\cite{chen-vassilicos-2022, apostolidis-laval-vassilicos-2023} for
non-homogeneous turbulent velocity fields, but here we apply it to the
turbulent scalar field. We therefore hypothesise that, to leading
order, $S_{T12}(r,y)$ is the same function of $r$ at different
positions $y$ as long as it is scaled by a characteristic temperature
$\tau(y)$, velocity $v(y)$ and a length $l_{T12}(Pr,y)$ that are
all local in $y$. We make the same hypothesis for $S_{T2}(r,y)$ and
allow for the possibility of a different length $l_{T2}(Pr,y)$:

\begin{gather}
S_{T2}(r,y) = \tau^2(y) a_{T2}(Pr) s_{T2}(r/l_{T2}(Pr,y),y^+ ) \label{eq:S_T2}\\ 
S_{T12}(r,y) = \tau(y) v(y) a_{T12}(Pr) s_{T12}(r/l_{T12}(Pr,y),y^+) \label{eq:S_T12} 
\end{gather}

{\flushleft where} the additional dependence on the local Reynolds number
$y^{+}$ accounts for higher order corrections to
our leading order hypothesis given that we consider asymptotics in the
limit $1\ll y^{+}\ll Re_{\tau}$ with $Pr = O(1)$ kept constant. A
dependence on the Prandtl number is retained in the form of
dimensionless functions $a_{T2} (Pr)$ and $a_{T12} (Pr)$ to be
determined.  Moreover, we define the non-dimensional function
$s_{T3}=S_{T3}/(\tau^2v)$.  Substituting into equation
(\ref{eq:budgt}) and using the high $Re_{\tau}$ approximation
$\varepsilon^v_T \approx u_\tau \theta_\tau^2/(k_\theta y)$  \citep{abe-antonia-2017}
in the intermediate/logarithmic layer,
we obtain:

\begin{gather}
k_\theta\frac{v\tau^2}{u_\tau \theta_\tau^2}\frac{s_{T3}}{r/y}
-\frac{3k_\theta y^2}{\pi r^3y^+} \frac{a_{T2}}{Pr} \int_0^{r/2} \rho^2
\frac{\text{d}^2 (\tau/\theta_\tau)^2s_{T2}}{\text{d}y^2} \,
\text{d} \rho  \nonumber \\- \frac{3k_\theta y^2}{\pi
  ry^+}\frac{a_{T2}}{Pr}\frac{\text{d}(\tau/\theta_\tau)^2s_{T2}}{\text{d}r}\approx
-1- a_{T12}\frac{6}{\pi r^3}\int_0^{r/2}\rho^2 \frac{\tau v}{\theta_\tau
  u_\tau}s_{T12} \, \text{d}\rho,
\label{eq:KHMHs}
\end{gather}

{\flushleft which} is the budget equation for the two-point passive scalar variance
in the intermediate region of a turbulent channel flow under the
hypothesis of a logarithmic layer for both the mean velocity and
scalar fields as might be expected (at least approximately) when
$Re_\tau \gg 1$ and $Re_\tau Pr \gg 1$. The hypothesis of homogeneous
physics in non-homogeneous turbulence is made both for the smaller
scales $r\ll l_{Tn,o}$ where $l_{Tn}$ are outer length scales
$l_{Tn,o}$ and for the larger scales $r\gg l_{Tn,i}$ where $l_{Tn}$
are inner length scales $l_{Tn,i}$, with $n=2$ and $12$.

\subsubsection{Outer similarity}

We start by obtaining from equation \ref{eq:KHMHs} the leading order
outer scale-by-scale budget equation. For these larger scales we
choose outer scalar, velocity and length scales
$\tau=\tau_{o}=\theta_\tau$, $v=v_{o}=u_{\tau}$ and
$l_{Tn,o}=l_{To}=y$. As a consequence, the outer form of equation
\ref{eq:KHMHs} is:

\begin{gather}
k_\theta\frac{s_{T3}}{r/y} -\frac{3k_\theta y^2}{\pi r^3y^+}
\frac{a_{T2}^o}{Pr} \int_0^{r/2} \rho^2 \frac{\text{d}^2 s_{T2}}{\text{d}y^2} \,
\text{d} \rho - \frac{3k_\theta y^2}{\pi
  ry^+}\frac{a_{T2}^o}{Pr}\frac{\text{d}s_{T2}}{\text{d}r} \nonumber\\ \approx-1-
\frac{6}{\pi r^3} a_{T12}^o\int_0^{r/2}\rho^2 s_{T12} \, \text{d}\rho.
\label{eq:KHMHso}
\end{gather}

This equation suggests outer asymptotic expansions of the form $s_{Tn}
= s_{Tn}^{o,0}+\frac{1}{y^+}s_{Tn}^{o,1}+...$ for $n = 2$, $12$ and $3$
in the limit $y^{+}\gg 1$, so that its leading order satisfies:

\begin{equation}
k_\theta\frac{s_{T3}^{o,0}}{r/y} \approx -1- \frac{6}{\pi
  r^3}a_{T12}^o\int_0^{r/2}\rho^2 s_{T12}^{o,0} \, \text{d}\rho.
\label{eq:KHMHso2}
\end{equation}

The outer two-point budget is a balance between normalised inter-scale
transfer rate on the left hand side and normalised dissipation ($-1$)
and production on the right hand side.

\subsubsection{Inner similarity}

We now obtain from equation \ref{eq:KHMHs} the leading order inner
scale-by-scale budget equation by concentrating on the physical
mechanisms at the smallest scales of the flow. For $r\ll
l_{Tn,o}(y)$, we set $\tau^2=\tau_{i}^{2}=\tau_{o}^2g^{2}(y^+)
=\theta_\tau^2g^{2}(y^+)$ and we define a length $l_{Ti}=l_{To}
f(y^{+})= yf(y^+)$ where $f$ and $g$ are a priori unknown decreasing
functions with increasing $y^+$. It is natural to assume that the
inner velocity scale $v_{i}$ is the same as the one obtained by
\cite{apostolidis-laval-vassilicos-2023} for the inner scaling of the
KHMH equation by a similar approach, and so we adopt
$v=v_{i}=u_\tau(1/y^+)^{1/4}$ which is the Kolmogorov inner velocity
scale $u_{\eta}\equiv (\nu \varepsilon^v)^{1/4}$. This accounts for
our limitation to the $Pr\le 1$ case for which there are no strong
sub-Kolmogorov scalar fluctuations. For the length scales $l_{Tn,i}$,
we assume a relation of the form $l_{Tn,i}=c_{Tn}(Pr)l_{Ti}$, where
$c_{Tn}$ are dimensionless functions of Prandtl number. Moreover, we
assume that the inner dependence on $Pr$ is uniquely represented by
the Prandtl relation of $l_{T2,i}$, hence $a_{T2}^i=a_{T12}^i=1$. As a
consequence, the inner form of equation \ref{eq:KHMHs} is:

\begin{gather}
k_\theta\Bigl(
\frac{1}{y^+}\Bigr)^{-1/4}\frac{g^2}{f}\frac{s_{T3}}{r/l_{Ti}}
-\frac{3k_\theta}{\pi r^3y^+} \frac{l_{T2,i}^2}{f^2} \frac{1}{c_{T2}^2Pr} \int_0^{r/2}
\rho^2 \frac{\text{d}^2 g^2s_{T2}}{\text{d}y^2} \, \text{d}\rho \nonumber
\\ -\frac{3k_\theta }{\pi
}\frac{1}{c_{T2}^2Pr}\Bigl(\frac{1}{y^+}\Bigr)\frac{g^2}{f^2}\frac{s'_{T2}}{r/l_{T2,i}}\approx
-1- \frac{6}{\pi r^3}\int_0^{r/2}\rho^2\Bigl(\frac{1}{y^+}\Bigr)^{1/4}g
s_{T12} \, \text{d}\rho
\label{eq:KHMHso3}
\end{gather}

{\flushleft where} $s'_{T2}=\text{d}s_{T2}/\text{d}(r/l_{T2,i})$. At inner scales the
leading order balance must involve inter-scale energy transfer,
molecular diffusion and dissipation without presence of production
(second term on the right hand side), which implies
$(\frac{1}{y^+})^{-1/4}\frac{g^2}{f}=(\frac{1}{y^+})\frac{g^2}{f^2}=O(1)$
independent of $y^+$. As a result, $g=(y^+)^{-1/4}$ and $f=
(y^+)^{-3/4}$, i.e. $\tau_{i}\sim \theta_{\tau}(y^{+})^{-1/4}\sim
\theta_\tau u_{\eta}/u_{\tau}$ and $l_{Ti}\sim \eta$ where $\eta\equiv
(\nu^{3}/\varepsilon^v)^{1/4}$ is the Kolmogorov length
scale.
From equation (\ref{eq:KHMHso3}), it follows that
$c_{T2}=1/\sqrt{Pr}$ so that $l_{T2,i} = c_{T2}(Pr) l_{Ti} \sim \eta/\sqrt{Pr}$
is the Batchelor length scale $\eta_B$ \citep{batchelor-1959} which
may be interpreted as the distance over which the scalar diffuses by
molecular action over a time proportional to the inverse
characteristic strain rate $\sqrt{\varepsilon^v/\nu}$. Note that our
analysis leads to the conclusion directly from the two-point budget
equation (without use of dimensional analysis) that the inner passive
temperature scale $\tau_{i}$ is $\theta_\tau u_{\eta}/u_{\tau}$, $l_{Ti}$
is $\eta$ and $l_{T2,i}$ is $\eta_B$.


As the budget equation must also balance beyond the leading order
where the production terms must feature, the inner asymptotic
expansions must be in terms of powers of
$(y^+)^{-1/4}g=(1/y^+)^{1/2}$, i.e.
$s_{Tn}=s_{Tn}^{i,0}+(\frac{1}{y^+})^{1/2}s_{Tn}^{i,1}+...$ for $n =
2$, $12$ and $3$ in the limit $y^{+}\gg 1$. The leading order inner
balance is:

\begin{equation}
k_\theta\frac{s_{T3}^{i,0}}{r/l_{Ti}} - \frac{3k_\theta }{2\pi
}\frac{s'^{i,0}_{T2}}{r/l_{T2,i}} \approx -1.
\label{eq:KHMHso4}
\end{equation}

\subsection{Matched asymptotics}
\label{sec:masym}

Matching the leading terms of the outer expansions for $r\gg l_{Tn,i}$ with the leading terms of the inner expansions for $r\ll y$
for $S_{T2}$ at fixed $y^+$ we obtain:
\begin{equation}
\underbrace{\tau_{o}^2 a_{T2}^o s^{o,0}_{T2}(r/y,y^+)}_{S^{o,0}_{T2}}=\underbrace{\tau_i^2 s_{T2}^{i,0}(r/\eta_B,y^+)}_{S_{T2}^{i,0}}.
\end{equation}

Using the result $g=(1/y^+)^{1/4}$:  

\begin{equation}
a_{T2}^o \Bigl(\frac{r}{y}\Bigr)^n=\Bigl(\frac{1}{y^+}\Bigr)^2 \Bigl(\frac{r}{\eta_B}\Bigr)^n
\end{equation}

{\flushleft leading} to $n=2/3$ and $a_{T2}^o=Pr^{1/3}$.
For $S_{T12}$, as at the largest scales the interscale transfer rate
approach zero, equation \ref{eq:KHMHso2} suggests $a_{T12}^o=1$.
By matching outer with inner expansions we obtain:

\begin{equation}
\underbrace{\tau_{o}v_o s^{o,0}_{T12}(r/y,y^+)}_{S_{T12}^{o,0}}=\underbrace{\tau_{i}v_i s^{i,0}_{T12}(r/l_{T12,i},y^+)}_{S_{T12}^{i,0}}, 
\end{equation}

{\flushleft leading to}:

\begin{equation}
\Bigl(\frac{r}{y}\Bigr)^n=\Bigl(\frac{1}{y^+}\Bigr)^2 \Bigl(\frac{r}{\eta c_{T12}}\Bigr)^n
\end{equation}

{\flushleft resulting} in $n=2/3$ and $c_{T12}^i=1$.  
The leading orders are $S^{0}_{T2}\sim
Pr^{1/3}(\varepsilon_{T}^{v}/\varepsilon^{v}) (\varepsilon^{v}
r)^{2/3}$ (similar to the Corrsin-Obukhov scalar spectrum for
homogeneous turbulence, see \cite{batchelor-1959}, but with an
additional $Pr^{1/3}$ prefactor) and $S^{0}_{T12}\sim
(\varepsilon_{T}^{v}/\varepsilon^{v})^{1/2} (\varepsilon^{v} r)^{2/3}$
in the intermediate range $y_{i} \ll r \ll y$. It also follows that
the inner scalings of $S_{T2}$ and $S_{T12}$ are, respectively,
$\theta_\tau^2 u_{\tau}^{-2} u_{\eta}^{2} F(r/l_{T2,i})$ and
$\theta_\tau u_{\tau}^{-1} u_{\eta}^{2} G(r/l_{Ti})$ ($F$, $G$
dimensionless functions).

The scalar inter-scale transfer rate $\varPi_{T}^{v}$ is obtained from
$S_{T3}$ for which we use the leading order outer and inner balances
(\ref{eq:KHMHso2}) and (\ref{eq:KHMHso4}). The former leads to
$S_{T3}^{o, 0} \approx -\varepsilon_{T}^{v} r [1 - A (r/y)^{2/3}]$
whilst the latter leads to $S_{T3}^{i,0} \approx -\varepsilon_{T}^{v}
r [1 - {B} (r/\eta_B)^{-4/3}]$ where $A$ and $B$ are dimensionless
constants. Following the rule of van Dyke \citep{vanDyke-1964}, the
composite leading order is $S_{T3}^{o, 0}$ plus $S_{T3}^{i,0}$ minus
their common part which is $-\varepsilon_{T}^{v}r$. Hence:
%
%
\begin{equation}
\varPi_T^v\approx - \varepsilon_T^v \Bigl(1-A(r/y)^{2/3}-\frac{B}{Pr^{2/3}}(r/\eta)^{-4/3}\Bigr). 
\label{eq:PITE}
\end{equation}

Equation \ref{eq:PITE} implies the existence of a scale $r=r_{min}$
where $\vert \varPi_T^v \vert/\varepsilon_T^v$ is maximal: $r_{min}$ and
the value $(\varPi_T^v/\varepsilon_T^v)_{min}$ of
$\varPi_T^v/\varepsilon_T^v$ at $r=r_{min}$ are ($\lambda$ being the
Taylor length)
\begin{equation}
r_{min}\sim \sqrt{\frac{\delta_\nu y}{Pr^{2/3}}}\sim
\frac{\lambda}{Pr^{1/3}}=\lambda_T
\label{eq:rmin1}
\end{equation}
\begin{equation}
1+(\varPi_T^v/\varepsilon_T^v)_{min} \sim (Pr^{2/3}y^+)^{-1/3} \sim Pr^{-2/9}Re_\lambda^{-2/3}.
\label{eq:rmin2}
\end{equation}
Similarly to the conclusion of
\cite{apostolidis-laval-vassilicos-2023} for the velocity field, in
the TCF's intermediate region, a Kolmogorov equilibrium
$\varPi_T^v=-\varepsilon_T^v$ can be reached asymptotically only
around $\lambda_T$, with systematic departures caused by
production at larger scales (term $A(r/y)^{2/3}$ in (\ref{eq:PITE}))
and by molecular diffusion at smaller scales (term
${B}(r/\eta_B)^{-4/3}$ in equation (\ref{eq:PITE})).

\subsection{Verification of the theory with the DNS data}
\label{sec:verification}

\begin{figure}
\includegraphics[width=\textwidth]{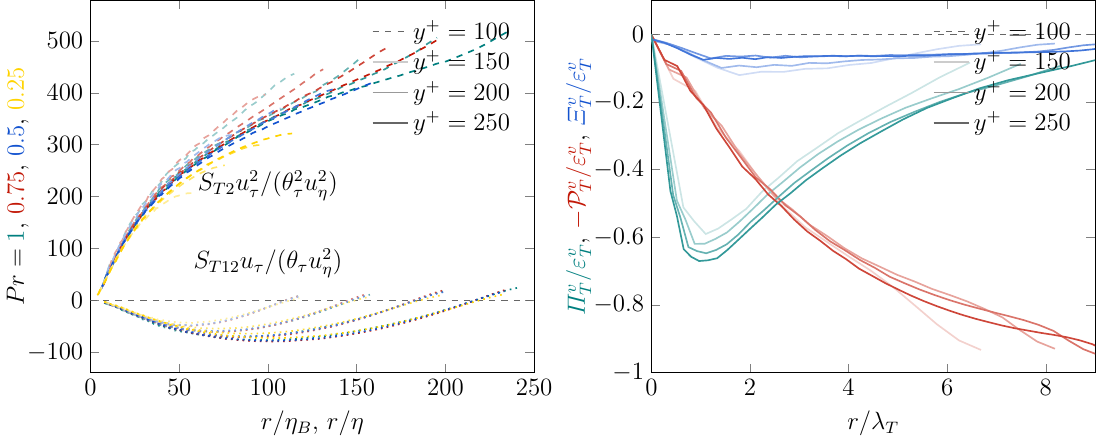}
\caption{Left panel: $S_{T2} u_\tau^2 /(\theta_\tau^2 u_\eta^2)$ and
  {$S_{T12} u_\tau /(\theta_\tau u_\eta^2)$} as functions of
  $r/\eta_B$ and $r/\eta$ respectively. Right panel:
  $\varPi^v_{T}/\varepsilon^v_T$ (green lines),
  $\mathcal{P}^v_T/\varepsilon^v_T$ (red lines), and transport terms
  $\varXi_T^v/\varepsilon^v_T=(\varPi_{Tm}^v+\mathcal{T}_{T}^v)/\varepsilon^v_T$
  (blue lines) as a function of $r/\lambda_T$ for $Pr=1$. Wall-normal
  distance is increased from light to dark colours.}
\label{fig:bal}
\end{figure}

In this section, we use TCF DNS data to verify the theoretical
conclusions in sections \ref{sec:asym} and \ref{sec:masym}. The top-left
panel of figure \ref{fig:bal} verifies the validity of the derived
inner scalings $S_{T2} = \theta_\tau^2 u_{\tau}^{-2} u_{\eta}^{2}
F(r/\eta_{B})$ and $S_{T12} = \theta_\tau u_{\tau}^{-1} u_{\eta}^{2}
G(r/\eta)$
by reporting how $S_{T2} u_\tau^2 /(\theta_\tau^2 u_\eta^2)$ and
{$S_{T12} u_\tau /(\theta_\tau u_\eta^2)$} evolve as functions of
$r/\eta_B$ and $r/\eta$ respectively. The collapse of the two sets of
curves at small scales is not perfect but satisfactory if we consider
that our DNS values of $Re_{\tau}$ and $Re_{\tau} Pr$ are not as large
as might be required by our asympotic analysis.

The right panel of figure \ref{fig:bal} represents different terms in
the balance equation \ref{eq:KHMHSTOT} integrated over the volume of a
sphere of radius $r/2$, at different wall-normal locations. The
results for the case $Pr=1$ are reported for conciseness, but similar
behaviour is observed for our other $Pr$ values. Green lines represent
inter-scale transfer rate and red lines represent production rate,
both normalised by the volume integral of the two-point dissipation
rate. In blue, we report the terms neglected in the budget equation
\ref{eq:budgt} to confirm that they are small: it is worth noticing
that their value is not zero, indicating that due to the limited value
of $Re_\tau$ these redistribution processes are low but still weakly
active.

The qualitative behavior of the different terms of the budget
equations is similar for the velocity
\citep{apostolidis-laval-vassilicos-2023} and the passive scalar.
The production rate increases as $r/\lambda_T$ increases above 1:
at the smallest scales below $\lambda_T$, the production rate is
small, but it is present and non-negligible at all scales larger than
$\lambda_T$ reaching, at the largest scales, the one-point
log-layer equilibrium $\mathcal{P}_T^v\approx\varepsilon_T^v$.

The inter-scale transfer rate is negative at all scales indicating a
cascade from large to small scales on average.
As predicted in the previous sub-section, the minima of
$\varPi_T^v/\varepsilon_T^v$ happen at similar values of
$r/\lambda_T$ for all $y^+$, and
$(\varPi_T^v/\varepsilon_T^v)_{min}$ decreases as $y^+$ and local
$Re_{\lambda}$ increase. The production-generated systematic departure
from Kolmogorov two-point equilibrium is present throughout the
inertial range, i.e. at all scales larger than $\lambda_T$.

\begin{figure}
\includegraphics[width=\textwidth]{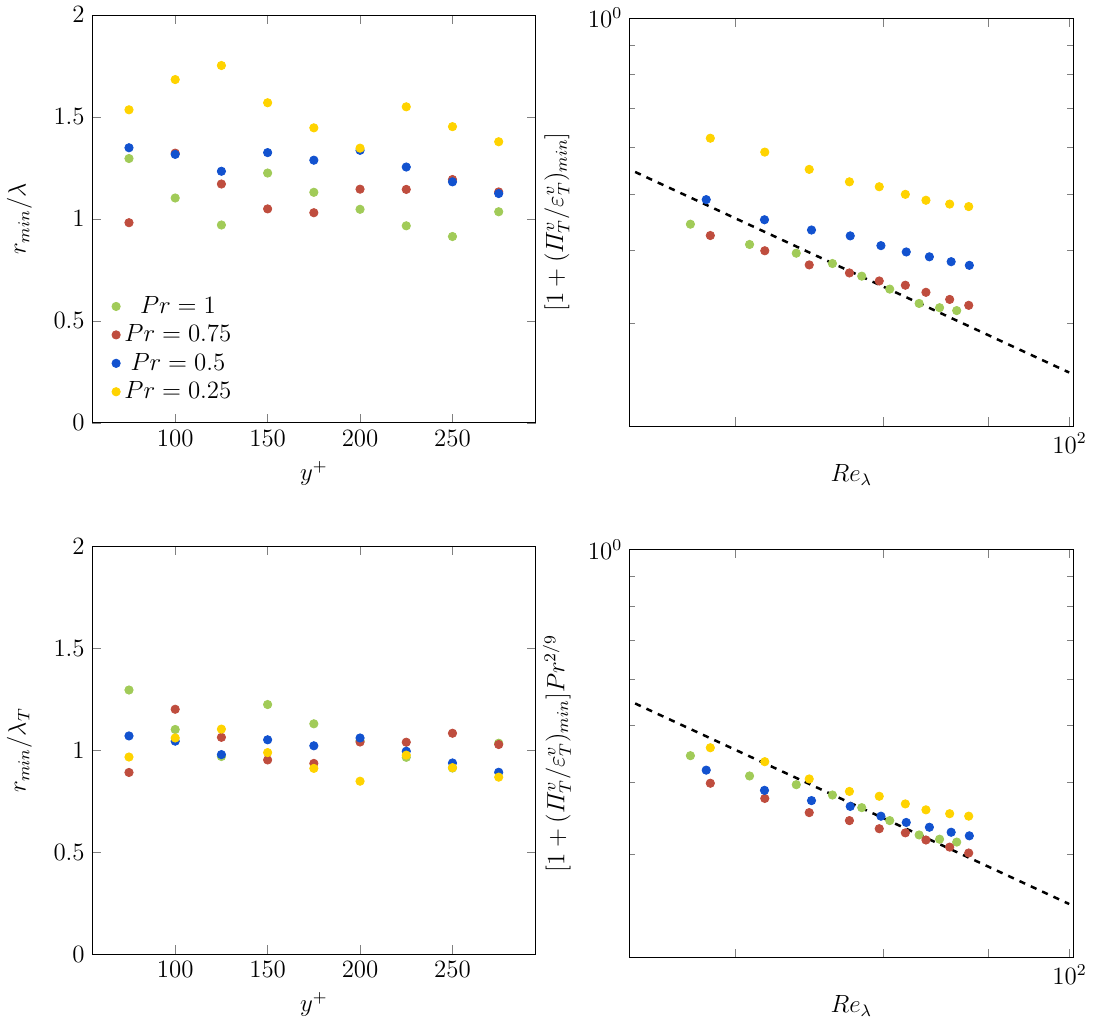}
\caption{Scale $r_{min}$ of $\varPi_T^v/\varepsilon_T^v$ minima as a
  function of wall distance $y^+$ (left) and values of $1 +
  \varPi_T^v/\varepsilon_T^v$ as a function of $Re_\lambda$ (right). Top
  panels: quantities not scaled with $Pr$. Bottom panels: quantities
  scaled with $Pr$. The dashed line in the right panels is proportional to $Re_\lambda^{-2/3}$.}
\label{fig:ver}
\end{figure}

The theoretical relations (\ref{eq:rmin1})-(\ref{eq:rmin2}) for the
passive scalar stemming from our matched asymptotic expansion are
quantitatively verified in figure \ref{fig:ver} for different $Pr$.
The left panels show the length-scale $r_{min}$ of the minimum of the
ratio between the inter-scale transfer and dissipation rates versus
$y^+$. In the top left panel, $r_{min}$ is normalised with $\lambda$,
in the bottom left panel with $\lambda_T$. For all cases, $r_{min}$
collapses once scaled with $Pr^{-1/3}$, i.e. as $r/\lambda_T$. The
minimum is at an almost constant value of $r/\lambda_T \sim 1$ for all
$y^+$ in the intermediate layer and all our $Pr$ values in agreement
with (\ref{eq:rmin1}).
The right panels of figure \ref{fig:ver} show
$1+(\varPi_T^v/\varepsilon_T^v)_{min}$ versus $Re_\lambda$ (which
varies with $y^+$). Considering the relatively low values of
$Re_{\lambda}$, the data collapse quite well on the line
$Re_\lambda^{-2/3}$ once multiplied by $Pr^{2/9}$ as per our
prediction (\ref{eq:rmin2}).  The turbulent Reynolds numbers
$Re_{\lambda}$ are not high, and their range is narrow in the
intermediate layer at $Re_{\tau}=1000$. Due to the limited Reynolds
number and the narrow dynamical range, deviations from the theoretical
model are to be expected. Similar deviations were also pointed out by
\cite{apostolidis-laval-vassilicos-2023} at $Re_\tau=1000$, who also
observed their reduction at $Re_\tau=2000$. The departure
from this theoretical prediction increases as $Pr$ decreases. Indeed,
the theoretical prediction is made for $Re_\tau \gg 1$ and $Re_{\tau}
Pr \gg 1$ and the condition $Re_{\tau} Pr \gg 1$ progressively weakens
with decreasing $Pr$ for a given $Re_{\tau}$ (as is the case in figure
2). The logarithmic region also becomes less well-defined as $Pr$
decreases \citep{alcantara-avila-etal-2018}. The best DNS agreement
with the theoretical scaling (\ref{eq:rmin2}) occurs for $Pr=1$ which
is the case where $RePr$ is the highest in our DNS data. As $Pr$
decreases while keeping $Re_{\tau}$ constant, the properties of the
scalar intermediate layer which lead to our theoretical predictions
deteriorate, namely, the assumed scaling of the turbulent scalar
dissipation rate $\varepsilon_{T}^{v}$
and the presence of a well-developed approximate log layer for the
passive scalar.  Finally, $Re_{\lambda}$ increases as $y^+$ increases
and as $y^+$ increases we move towards the edge and out of the
intermediate layer where the theoretical predictions are designed for.


\subsection{Inter-scale transfer rate analysis}

\begin{figure}
\includegraphics[width=\textwidth]{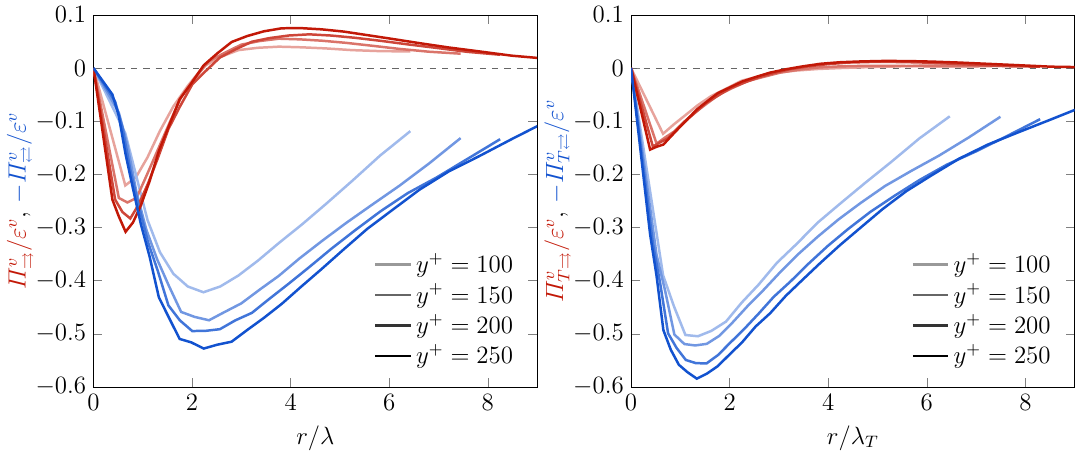}
\caption{Decomposition of the inter-scale transfer rate into
  anti-aligned (blue lines) and aligned (red lines)
  contributions for $Pr=1$. Left panel: Velocity; Right panel: Temperature.}
\label{fig:intA-AA}
\end{figure}

Given that the previous sub-section's results for the passive scalar
flucutations are similar to those of
\cite{apostolidis-laval-vassilicos-2023} for the velocity fluctuations
in the same flow, this section focuses on a more detailed comparison
between inter-scale transfer of turbulent energy and turbulent scalar
variance.
In figure \ref{fig:intA-AA} we concentrate on our highest $Re_{\tau}
Pr$ value (i.e. $Pr=1$) and plot inter-scale transfer rates decomposed
into aligned and anti-aligned contributions for the velocity (left
panels) and temperature (right panels). Aligned and anti-aligned
contributions arise from the following decomposition introduced by
\cite{apostolidis-laval-vassilicos-2023} for the inter-scale turbulent
energy transfer rate $\varPi^v$ and explicitely given here for $\varPi_T^v$:
\begin{equation} 
\varPi_T^v=\varPi_{T\rightleftarrows}^v+\varPi_{T\rightrightarrows}^v = \frac{3}{4\pi}\int \text{d}\Omega \aver{ \frac{\delta \bm{u}'\cdot \hat{\bm{r}}}{r} \delta \theta'\delta \theta'}_\rightrightarrows+\frac{3}{4\pi}\int \text{d}\Omega \aver{ \frac{\delta \bm{u}'\cdot \hat{\bm{r}}}{r} \delta \theta'\delta \theta'}_{\rightleftarrows}
\end{equation}
where the conditional average $\aver{...}_\rightrightarrows$ is taken
over pairs of points where fluctuating velocities are aligned,
i.e. $\bm{u}'(\bm{x}+\bm{r}/2)\cdot\bm{u}'(\bm{x}-\bm{r}/2)>0$,
and the conditional average $\aver{...}_{\rightleftarrows}$ is taken
over pairs of points where fluctuating velocities are anti-aligned,
i.e.  $\bm{u}'(\bm{x}+\bm{r}/2)\cdot\bm{u}'(\bm{x}-\bm{r}/2)<0$.
Both $\varPi^v$ and $\varPi_T^v$ involve the multiplication of the
second-order structure functions (respectively of velocity and
temperature) and the projected velocity difference $\delta \bm{u}'
\cdot \hat{\bm{r}}$. The local sign of the velocity and temperature
cascades is fully determined by $\delta \bm{u}' \cdot \hat{\bm{r}}$ ,
with compressions ($\delta \bm{u}' \cdot \hat{\bm{r}}<0$) causing
forward cascades and stretchings ($\delta \bm{u}' \cdot
\hat{\bm{r}}>0$) causing inverse cascades. Clearly, the local sign of
the cascade is not enough to compute its average value which stems
from weighted averaging as per the above equation. Note that both
aligned and anti-aligned components can involve both compressions and
stretchings and therefore result in either a forward or inverse
cascade component.  Our results for the scalar fluctuations and those
of \cite{apostolidis-laval-vassilicos-2023} for the velocity
fluctuations show that $\varPi_T^v$ and $\varPi^v$ have similar evolutions
with $r/\lambda_T$ and $r/\lambda$ respectively, and that they
are both negative at all the scales, meaning that local compressions
dominate over local stretchings on average. In figure
\ref{fig:intA-AA} we compare how aligned and anti-aligned pairs of
velocity contribute to the cascade for the velocity and the cascade
for the scalar. The anti-aligned components decrease and reach a
minimum at a scale $r>r_{min}$,
and then increase towards zero as expected since production balances
dissipation at the largest scales. They dominate the inter-scale
transfer rates at scales $r> r_{min}$ and show similar evolutions with
$r/\lambda$-$r/\lambda_T$ for $\delta \bm{u}'\delta \bm{u}'$ and
$\delta \theta'\delta \theta'$. The aligned contributions to the
inter-scale transfer rate also show similar evolutions with
$r/\lambda$-$r/\lambda_T$ for $\delta \bm{u}'\delta \bm{u}'$ and
$\delta \theta'\delta \theta'$, reaching a minimum at a scale
$r<r_{min}$ for both the scalar and the velocity cases.
Despite these similar evolutions, the aligned contributions are
clearly different as their magnitude is significantly lower for the
temperature than for the velocity fields. Furthermore, the peak value
of $\varPi_{\rightleftarrows}^v/\varepsilon^v$ appears to coincide in
$r$ with the zero value of
$\varPi_{\rightrightarrows}^v/\varepsilon^v$ (where $\varepsilon^v$ is
the sphere-averaged turbulence dissipation rate). If this coincidence
(\ref{fig:intA-AA}) persists as $Re_{\tau}\to \infty$ then the
intermediate asymptotic theory of
\cite{apostolidis-laval-vassilicos-2023} would imply that
$\varPi_{\rightleftarrows}^v/\varepsilon^v$ tends to $-1$ in that
limit. However, the peak value of
$\varPi_{T\rightleftarrows}^v/\varepsilon_{T}^v$ does not coincide in
$r$ with the zero value of
$\varPi_{T\rightrightarrows}^v/\varepsilon_{T}^v$ as clearly seen in
\ref{fig:intA-AA}. If this lack of coincidence persists as
$Re_{\tau}\to \infty$ then (\ref{eq:rmin2}) would imply that
$\varPi_{T\rightleftarrows}^v/\varepsilon_{T}^v$ does not tend to $-1$
in that limit. Even though the overall physical process determining
the direction of the cascade is the same for velocity and temperature
(compressions and stretchings via $\delta \bm{u}' \cdot \bm{n}$), there
are significant differences arising from whether compressions and
stretchings result from aligned or anti-aligned fluctuating
velocities. The explanation of these differences will require a
dedicated future study which may need to consider the underlying flow
structure of the turbulence, for example, in terms of ejections and
sweeps.



%% file: tex/Conclusions.tex
\section{Conclusions}
\label{sec:conclusions}

We performed a scale-by-scale analysis of passive scalar fluctuations
in the intermediate layer of a fully developed turbulent channel flow
to assess the presence or absence of a Kolmogorov-like
equilibrium. The turbulent channel flow analysis of
\cite{apostolidis-laval-vassilicos-2023} has had to be non-trivially
extended to account for a passive scalar with Prandtl number $Pr \leq
1$ in the limit $Re_{\tau} Pr \gg 1$.
It transpired from our hypothesis of homogeneous physics in
non-homogeneous turbulence combined with the Navier-Stokes equation
that the scalar fluctuations' inner length scale is the Batchelor
length even though the turbulence in the channel flow's intermediate
layer is not homogeneous down to very small scales.  Indeed, the
Batchelor length was originally introduced by \cite{batchelor-1959}
for homogeneous turbulence, yet two-point turbulent scalar production,
and therefore non-homogeneity, in the channel flow's intermediate
layer are significant down to length-scales as small as
$\lambda_{T} = \lambda Pr^{-1/3}$.
It also follows from our analysis that the scalar field's average interscale
transfer rate is negative (forward cascade on average) and maximal at
a specific length-scale $r_{min}\sim \lambda_{T}$. The tendency
towards a Kolmogorov equilibrium between scalar interscale transfer
rate and scalar dissipation rate is achieved asymptotically
($Re_{\tau} \to \infty$) at that length scale only, and systematic
departures from this equilibrium are present and increase with
length-scale throughout the inertial range because of the presence of
two-point scalar turbulence production.
These theoretical arguments have been verified with DNS of turbulent
velocity and scalar fields in channel flow at different Prandtl
numbers $Pr=1$, $0.75$, $0.5$, and $0.25$. Good agreement has been
found between the hypotheses and the predictions on the one hand, and
the DNS data on the other. Even though the aforementioned results and
respective ones in \cite{apostolidis-laval-vassilicos-2023}
demonstrate a close analogy between the interscale transfer properties
of turbulent velocity and scalar fields in the intermediate layer of
turbulent channel flow, there are some significant differences between
them. Whilst the interscale transfer rates $\varPi^v$ (for the
fluctuating velocity) and $\varPi_T^v$ (for the fluctuating scalar) have
similar evolutions with length-scale $r$, 
the contributions from aligned and anti-aligned pairs of fluctuating
velocities differ for $\varPi^v$ and $\varPi_T^v$.